\documentclass[aps,prb,twocolumn,groupedaddress]{revtex4}

\usepackage{amsmath,amssymb}
\usepackage{graphicx}
\usepackage{epstopdf}
\usepackage{bm}
\usepackage{color}

\newcommand{\slr}{$(T_1T)^{-1}$}
\newcommand{\cfca}{Ca(Fe$_{1-x}$Co$_x$)$_2$As$_2$}
\newcommand{\bfca}{Ba(Fe$_{1-x}$Co$_x$)$_2$As$_2$}

\begin{document}


\title{Pseudogap-like phase in \cfca\ revealed by $^{75}$As NQR}


\author{S.-H. Baek}
\email[]{sbaek.fu@gmail.com}
\affiliation{IFW-Dresden, Institute for Solid State Research,
PF 270116, 01171 Dresden, Germany}
\author{H.-J. Grafe}
\affiliation{IFW-Dresden, Institute for Solid State Research,
PF 270116, 01171 Dresden, Germany}
\author{L. Harnagea}
\affiliation{IFW-Dresden, Institute for Solid State Research,
PF 270116, 01171 Dresden, Germany}
\author{S. Singh}
\affiliation{IFW-Dresden, Institute for Solid State Research,
PF 270116, 01171 Dresden, Germany}
\author{S. Wurmehl}
\affiliation{IFW-Dresden, Institute for Solid State Research,
PF 270116, 01171 Dresden, Germany}
\author{B. B\"{u}chner}
\affiliation{IFW-Dresden, Institute for Solid State Research,
PF 270116, 01171 Dresden, Germany}
\date{\today}

\begin{abstract}
We report $^{75}$As NQR measurements on single crystalline \cfca\
($0\leq x \leq 0.09$). The nuclear spin-lattice relaxation rate $T_1^{-1}$ as a
function of temperature $T$ and Co dopant concentration $x$ reveals a gradual
decrease of \slr\ below a
crossover temperature $T^*$ in the under- and optimally-doped region. Since
there is no hint of a thermodynamic phase
transition near $T^*$, this spin-gap behavior is attributed to the
presence of a pseudogap-like novel state of electrons above $T_c$. The
resulting $x$-$T$ phase diagram shows that, after suppression of the
spin-density-wave order, $T^*$ intersects $T_c$ falling to zero rapidly near the optimal
doping regime.  Possible origins of the pseudogap behavior are discussed.
\end{abstract}




\maketitle

\section{Introduction}

The newly discovered iron-based pnictide superconductors (iron
pnictides) show some striking similarities with the cuprates. They
are composed of a layered structure with electronically-active
planes containing Fe or Cu, respectively. Superconductivity arises
from magnetically ordered parents by suppressing the magnetism
through chemical doping or pressure.\cite{lumsden10a} Despite a
large effort in recent years, there has been no conclusive answer
as to whether the iron pnictides share the same underlying physics
with the cuprates. Beyond a probably different superconducting gap
symmetry,\cite{mazin08a,fletcher09,kontani10} an important
difference is that the parent compounds of iron pnictides are
itinerant. Electronic correlations are less important and there is
no obvious link to Mott physics.\cite{koitzsch08,kroll08,hansmann10} So far, there is no
consensus about the precise nature of the normal state in iron
pnictides, which is crucial to understand the high temperature
superconductivity. A key question is whether there exists an
exotic state of matter in the normal state, equivalent to the
pseudogap phase in cuprates,\cite{timusk99,norman05} and where it
originates from.\cite{ikeda08,lee10,ishida10a} In particular, in
iron pnictides there are inconsistencies with respect to what can
rightfully be called a pseudogap behavior, with contradicting
experimental evidence on its presence itself, on its doping
dependence, and on the region of the phase diagram where it would
occur.\cite{sato08,hess09,mertelj09,kwon10,massee10,klingeler10}

While the nuclear spin-lattice relaxation rate $T_1^{-1}$ has proven to be an
excellent probe of the pseudo spin-gap phase in most
cuprates,\cite{warren89,ishida98a} it has not been successful in identifying
such a
novel phase, particularly, in the underdoped region of the iron pnictides.
Instead, strong antiferromagnetic (AFM) spin fluctuations
(SF) above the magnetic order exist in undoped compounds, and
persist even up to optimal doping in the superconducting samples,\cite{nakai08a,
ning09, ning10, lang10, dioguardi10} leading to a boost of \slr\ above the
magnetic or superconducting transition temperature.
Uniquely among iron pnictides, the parent compound CaFe$_2$As$_2$
does not show such strong SF above the magnetic and structural
transition temperatures $T_N=T_S$.\cite{baek09} Thus, CaFe$_2$As$_2$
could be an ideal system to search the pseudogap-like phase which may arise
from the weak electron correlations.

In this paper, we present a
systematic $^{75}$As NQR study of \cfca\ ($0\leq x \leq 0.09$)
focusing on the doping dependent normal state properties. In the
under- and optimally-doped regions, measurements of $T_1^{-1}$
show an anomalous suppression in the spectral weight of the
low-energy spin dynamics below a crossover temperature $T^*$, i.e.
pseudogap-like behavior. In contrast to other iron pnictides,
$T^*$ shows a strong doping-dependence, falling to zero near
optimal doping.

\section{Sample preparation and experimental details}

Single crystals of \cfca\ were grown in Sn flux and their basic
physical properties have been well
characterized.\cite{klingeler10, pramanik10, harnagea11} $^{75}$As
($I=3/2$) nuclear quadrupole resonance (NQR) measurements were
carried out in 5 different compositions of \cfca\ in the range
$0\leq x \leq 0.09$, where the Co concentration $x$ was 
determined by energy dispersive x-ray 
(EDX) analysis. The samples measured in our study are identical to those 
used in Ref. \onlinecite{harnagea11} and therefore we used throughout this 
paper the SDW transition 
$T_N$, the structural transition $T_S$,  
and the superconducting  
transition $T_c$ obtained in Ref.~\onlinecite{harnagea11}. We also confirmed 
the occurrence of AFM order at $T_N$ from the NQR spectrum which is   
considerably broadened below $\sim T_N$ due to the distribution of the local 
field at the nuclear sites, suggesting incommensurate magnetic order. 

Often, NQR is advantageous over NMR since it 
does not require an external field that may induce
additional magnetic effects. In fact, the single crystals of \cfca\
are extremely soft so that an external field can cause very
inhomogeneous NMR broadening, which strongly varies depending on
each piece of crystal, making NMR rather inappropriate to study a
systematic doping dependence. The material, however, features a
large $^{75}$As nuclear quadrupole frequency $\nu_Q$, which allows
NQR without the need for an external magnetic field.

$T_1^{-1}$ was obtained by
the saturation recovery method, where the relaxation of the nuclear
magnetization after a saturating pulse was fitted with a single exponential
function,
\begin{equation}
\label{}
1-M(t)/M(\infty)=a\exp(-3t/T_1),
\end{equation}
where $M$ is the nuclear magnetization and $a$ a fitting parameter
that is ideally one.  Since the total linewidth of the NQR
spectrum is somewhat broader than the bandwidth of the NQR
resonant circuit, in particular for larger doping as shown in
Fig. 1, we carefully measured $T_1^{-1}$ in order to avoid any
artifact such as spectral diffusion. In most cases, however,
the data are well fitted using the above function, as
shown in Fig. 2 for the optimally doped compound. Also, we
confirmed that $T_1^{-1}$ is unique over the whole spectrum. 

\section{Results and discussion}

Fig. 1 shows the $^{75}$As NQR spectra of \cfca\ for different Co
concentrations $x$ at room temperature and their fits to Gaussian
lines.\cite{footnoteshoulder_cfca} The linewidth of the spectrum
increases with increasing $x$, as expected from a progressively
increasing inhomogeneous distribution of the electric field
gradient (EFG). The linewidth of 400 kHz at $x=0$ increases up to
950 kHz at $x=0.09$. Interestingly, the resonance frequency, which
is the same as $\nu_Q$ for $I=3/2$ in the axial symmetry, slightly
increases with increasing $x$. Similar behavior was also reported
in $R$FeAs(O$_{1-x}$F$_x$) ($R$=La,Sm).\cite{lang10} The origin
of the behavior of $\nu_Q(x)$ is unknown, but the multi-orbital
electronic states of Fe$^{2+}$ ($d^6$) and considerable overlap
with $p$-orbitals of the As ion could be very sensitive to
dopants, giving rise to the increase of the EFG at the $^{75}$As.
A detailed analysis of the EFG and its temperature dependence in \cfca\ will
be reported elsewhere.

\begin{figure}
\centering
\includegraphics[width=\linewidth]{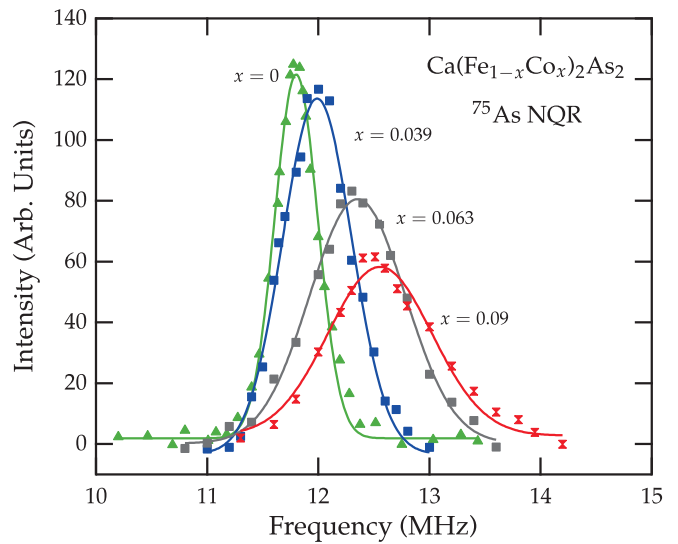}
\caption{\label{fig:spec} $^{75}$As NQR spectra in \cfca\ ($0\leq x \leq 0.09$)
at room temperature.
Both resonance frequency $\nu_Q$ and linewidth increase with
increasing $x$.}
\end{figure}

In Fig. 3 (a) and (b), $T_1^{-1}$ and $(T_1T)^{-1}$, respectively,
measured at the peak of the $^{75}$As NQR spectra, are shown as a
function of temperature and $x$. Above 200 K, \slr\ is independent
of temperature, and insensitive to $x$ for all compositions
including the undoped compound. While this is in contrast to other
pnictides where \slr\ is reduced with increasing doping, and still
shows a temperature dependence, the constant behavior of \slr\ in
\cfca\ indicates a weakening of correlations and a convergence to
Fermi liquid (FL) behavior at high temperatures. Thus, the unique
behavior of \slr\ in \cfca\ among iron pnictides indicates a
substantial weakening of AFM SF due to the strongly first order
SDW transition whose nature is well preserved against doping. It
is also noticeable that $(T_1T)^{-1}= 0.97$ s$^{-1}$K$^{-1}$ in
\cfca\ is much larger than the values in other iron pnictides
e.g., an order of magnitude larger\cite{nakai08a,lang10} than in
1111 and a factor of three larger even for the isostructural
BaFe$_2$As$_2$.\cite{baek08a,ning09} The large value \slr\ is
ascribed to the larger spectral density of spin fluctuations
as compared to other iron pnictides.\cite{baek09} Together
with the largest $\nu_Q$ among 1111 and 122 materials, it appears
that the material is the most extreme iron pnictide regarding the
sensitivity to the out-of-plane structure.

\begin{figure}
\centering
\includegraphics[width=\linewidth]{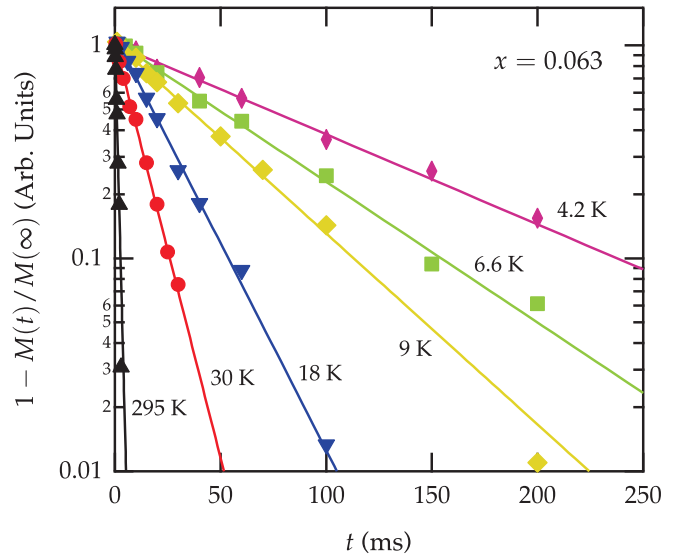}
\caption{\label{fig:recov} Recovery of the $^{75}$As nuclear
magnetization $M(t)$ as a
function of time $t$ for optimally doped \cfca\ ($x=0.063$).
Single exponential function (solid lines) fits data quite well in the whole temperature
range investigated. }
\end{figure}

In the parent compound ($x=0$), weak but distinct enhancement of \slr\ is
visible below 200 K down to $T_N$.
In \cfca, the EFG at the $^{75}$As is directed along the $c$ axis, quantizing
nuclei along $c$. Thus, the relaxation of the $^{75}$As on the NQR is 
expected to be
the same as that on the NMR for $H\parallel c$, unless the relaxation 
mechanism is different for the two cases. Indeed, the \slr\ data
measured by  NQR are in
excellent agreement with previous $^{75}$As NMR results\cite{baek09} in the
external field $H \parallel c$, confirming that the weak
enhancement of \slr\ above $T_N$ reflects intrinsic low-energy spin dynamics
in the material, as well as demonstrating good experimental resolution.
At a doping level of $x=0.039$, \slr\ replicates the data of the
parent compound with decreasing temperature. Then, soon after passing
$T_N(x=0)$, an unexpected suppression of \slr\ is observed at a temperature
higher than $T_N(x=0.039)$. Here, the temperature below which \slr\ shows an
anomalous suppression is defined as a crossover temperature $T^*$. With
further increasing $x$ to $0.056$, $T^*$ is slightly reduced, while $T_N$ is
markedly suppressed. At the optimal doping of $x=0.063$, the SDW is
completely suppressed, but the maximum in \slr\ is still clearly visible
and $T^*$ is further reduced. The peak of \slr\ centered at $T^*$ broadens
with increasing $x$ up to the optimal doping, and disappears in the
overdoped region ($x=0.09$).

Generally, \slr\ can be expressed in terms of the dynamical
spin susceptibility
$\chi''(\mathbf{q},\omega_0)$,\cite{moriya63}
\begin{equation}
\label{}
 (T_1T)^{-1} \propto \gamma_n^2 \sum_\mathbf{q} A^2(\mathbf{q})\chi''(\mathbf{q},\omega_0),
\end{equation}
where $\gamma_n$ is the nuclear gyromagnetic ratio, $A(\mathbf{q})$ the
$\mathbf{q}$-dependent hyperfine coupling constant, and
$\chi''(\mathbf{q},\omega_0)$ the imaginary
part of the dynamical
susceptibility at the Larmor frequency $\omega_0$. Since $A$ is usually
temperature independent, the suppression of \slr\ indicates that the low
energy dynamical susceptibility decreases below $T^*$, exhibiting 
spin gap-like behavior.
Clearly, $T^*$ is well separated from $T_N$ as well as $T_c$, as shown
in Fig. 4.\cite{harnagea11} We emphasize that $T^*$ is not associated with the structural
transition at $T_S$ either. Although $T_S$ seems to be decoupled from
$T_N$ with doping,\cite{harnagea11} $T_S$ occurs slightly above $T_N$ but
well below $T^*$. In our NQR study, however, we could not observe any 
visible change of the NQR frequency $\nu_Q$ or EFG,
which is a good probe of the structural transition, near $T^*$ even down to $T_N$. 
Furthermore, at optimal doping, both $T_N$ and $T_S$ 
disappear,\cite{harnagea11} while $T^*$ is clearly observed (see Fig. 4).
Therefore, $T^*$ is interpreted as a crossover temperature below which the
pseudogap-like phase emerges. 

\begin{figure}
\centering
\includegraphics[width=\linewidth]{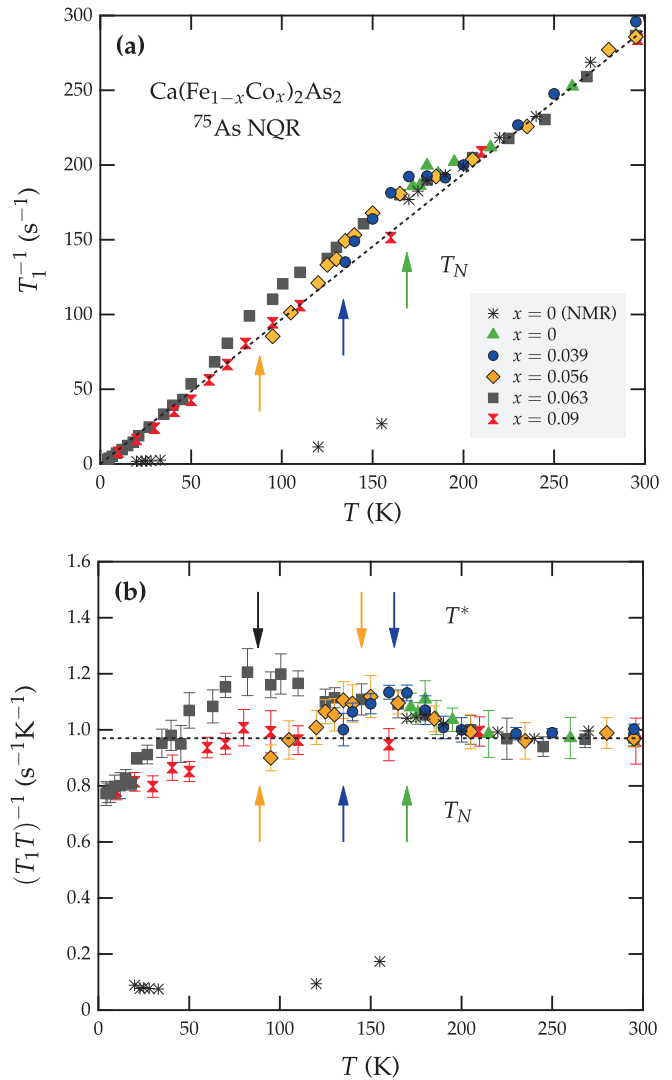}
\caption{\label{fig:T1T} Nuclear spin-lattice relaxation rate $T_1^{-1}$
versus temperature. Up arrows denote the SDW transition temperature $T_N$. $T_N$ is
not observed at optimal $x=0.063$ and above. }
\end{figure}

From these observations, together with $T_N$ and $T_c$ extracted from the
uniform magnetic susceptibility measurements,\cite{harnagea11} we draw the phase diagram of
\cfca\ in Fig. 4. $T_c$ is depicted together with the
superconducting (SC) volume fraction, which was estimated from the
diamagnetic response of the susceptibility in zero field cooled
measurements.\cite{harnagea11} It is noteworthy that ``bulk''
superconductivity with 100 \% SC
volume is achieved only in the vicinity of optimal-doping and the SC volume
fraction diminishes rapidly with underdoping or overdoping. For example,
for $x=0.039$, only 10 \% of the sample volume is estimated to be
superconducting. This trend rules out the role of chemical inhomogeneities
for the sizable change of the SC volume fraction. Rather, it is likely the
result of the competition between the SDW and the SC phase.   In support of
this, the SC volume seems to increase linearly as
$T_N$ is reduced reaching 100 \%
immediately after $T_N$ is completely suppressed.

\begin{figure}
\centering
\includegraphics[width=\linewidth]{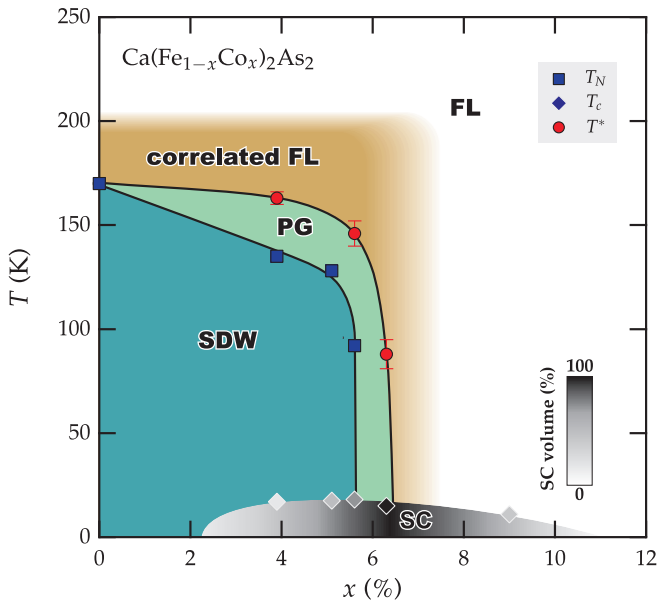}
\caption{\label{fig:phase}
$T$-$x$ phase diagram in \cfca\ obtained by
\slr\ measurements on the $^{75}$As NQR and by the uniform magnetic
susceptibility $\chi_\text{DC}$. FL and PG represent Fermi liquid and
pseudogap, respectively. Superconducting volume fraction was estimated by the
diamagnetic response of $\chi_\text{DC}$. }
\end{figure}

Up to now, pseudogap behavior in other iron pnictides has been
inferred from a decrease of the static susceptibility\cite{nakai08a,
klingeler10, ning09} (Knight shift and
macroscopic susceptibility measured by SQUID)
 or from a concomitant decrease of
\slr\  with decreasing temperature at high temperatures.\cite{grafe08,ning09}
Even in \cfca, the static susceptibility
shows a similar behavior.\cite{klingeler10}
However, a doping dependence of
such a pseudogap behavior has never been observed, lacking a relationship to
superconductivity as well as to the SDW phase at lower doping levels.
In contrast, our data reveal the strong doping dependence of $T^*$ in the
under- and optimally-doped region as shown in the phase diagram (Fig. 4),
which is analogous to that observed in cuprates.
Furthermore, we could not observe any anomaly at $T_c$ for optimal doping,
which is also similar to
underdoped cuprates.\cite{ishida98a} Nevertheless, the weak temperature
dependence of $T_1^{-1}$ in the superconducting state is unexpected, and has
only been observed in heavily overdoped Ba$_{1-x}$K$_x$Fe$_2$As$_2$.\cite{zhang10}

Besides, the suggested pseudogap size is unlikely large, being inconsistent with
its very weak effect demonstrated in our study.
In the isostructural Ba analog \bfca,\cite{ning09,ning10} for example, a
doping independent gap magnitude of $\Delta_\text{PG}\gtrsim 450$ K has been
estimated from the Knight shift and \slr, using the
phenomenological pseudogap function,
$(T_1T)^{-1} \text{ or } \mathcal{K} \propto \exp(-\Delta_\text{PG}/T)$. In
comparison, in \cfca, the
similar fit to \slr\ data
below $T^*$ for optimal $x=0.063$ [Fig. 3(b)] gives rise to
$\Delta_\text{PG}\sim 50$ K, which is an order
of magnitude smaller than the values inferred in the Ba-counterpart.
Then, the pseudogap behavior in other iron pnictides may be outweighed by AFM
correlations that cause an upturn of \slr. A conservation of the first order
character of the SDW transition\cite{cano10} in \cfca\ may inhibit strong AFM
correlations, allowing the observation of the pseudogap behavior.  Note that
also in some
cuprates as, for example, La$_{2-x}$Sr$_{x}$CuO$_4$, an upturn of \slr\ is
observed instead of a pseudogap, which is presumably due to short-range AFM
correlations.

Now we discuss the possible origin of the pseudogap-like phase in
\cfca.  In the phase diagram, $T^*$ appears to merge with $T_N$ as
$x\rightarrow 0$, and falls steeply to zero intersecting the SC
dome near optimal doping (100 \% SC volume fraction) where the
long range SDW order disappears. $T^*\sim T_N$ at $x=0$ and the
similar doping dependences of both $T^*$ and $T_N$ suggest that
the pseudogap and the SDW may originate from the same physical
basis. Also, the existence of $T^*$ at $x=0$ suggests the itinerant
origin of the pseudogap.  In comparison, in undoped cuprates i.e.,
Mott insulators, the pseudogap is always absent. Indeed, a recent
NMR study of Bi$_2$Sr$_{2-x}$La$_x$CuO$_{6+\delta}$ reveals that
the pseudogap ground state is metallic and that it is suppressed
abruptly near the antiferromagnetic ordered, Mott insulating
phase.\cite{kawasaki10a} One may speculate that the pseudogap emerges
from the \textit{incomplete} nesting of the Fermi surface, which
is not associated with the SDW order. Since doping would degrade
the nesting condition progressively, both $T^*$ and $T_N$ are
decoupled with increasing doping.

From the seeming correlation between $T^*$ and $T_c$ near optimal
doping in the phase diagram and from the incomplete nesting
scenario as an origin of the pseudogap, one may conjecture an
intriguing scenario that the pseudogap is an \textit{incoherent}
spin-gapped phase as a precursor state for the coherent SC phase.
Recent Andreev reflection studies in \bfca\ reveal
phase-incoherent SC pairs\cite{sheet10} above $T_c$  which may
support this scenario. The weak superconductivity (20\% SC volume)
in the overdoped region ($x=0.09$) can be also explained in this
scenario since a drop of \slr\ at 60 K [Fig. 3(b)] may indicate the
remaining pseudogap that coexists with the dominating FL phase.


\section{Conclusion}

In conclusion, we have measured the spin lattice relaxation rate, $T_1^{-1}$,
by means of $^{75}$As NQR in \cfca.   At
high temperatures, \slr\ is
independent of doping and
temperature. With lowering temperature, \slr\ as a function of $x$ and $T$
reveals a pseudogap behavior.

In contrast to other iron pnictides, the pseudogap in \cfca\ is strongly doping 
dependent, and appears to be an order of magnitude smaller.  
From the doping dependence shown in the $x$-$T$ phase diagram, we interpret that the 
pseudogap is strongly related to the SDW 
ordered phase at low doping, and could be a precursor state for
the coherent SC phase. 
Moreover, the $T$ dependence of the static susceptibility
(Knight shift) is other than that of the spin lattice
relaxation, which puts constraints on theoretical descriptions of the
pseudogap behavior in iron pnictides.

\section*{Acknowledgment}

We thank M. Deutschmann, S. Pichl, and J. Werner for experimental
help, and G. Lang for discussion. This work has been supported by
the Deutsche Forschungsgemeinschaft through SPP1458 (Grants No.
GR3330/2 and No. BE1749/13).

\bibliography{mybib}

\begin{thebibliography}{36}
\bibitem[{foo()}]{footnoteshoulder_cfca}
\bibinfo{note}{For optimal- and overdoped samples, a shoulder appears on the
  high frequency side of the spectra below $\sim 90$ K ($x = 0.063$) and below
  $\sim 200$ K ($x$ = 0.09). For the underdoped sample ($x = 0.039$) such a
  shoulder does not appear above the structural transition temperature,
  $T_S\sim 140$ K.\cite{harnagea11} Since $T_1^{-1}$ at the shoulder for
  optimal doping is confirmed to be the same as the one at the main peak, we
  conclude that the shoulder arises from a slightly different environment such
  as the collapsed tetragonal phase.}
\end{thebibliography}

\end{document}